\documentclass[draft]{agujournal2019_arxiv}
\usepackage{url} 
\usepackage{amsmath}
\usepackage{siunitx}
\usepackage{soul}
\usepackage{multicol}
\usepackage{xcolor}

\draftfalse

\journalname{Geophysical Research Letters}

\begin{document}
\justifying
%
%

\title{Reanalysis-based Global Radiative Response to Sea Surface Temperature Patterns:\\Evaluating the Ai2 Climate Emulator}

%
%

\authors{Senne Van Loon, Maria Rugenstein, \& Elizabeth A. Barnes}
\affiliation{}{Department of Atmospheric Science, Colorado State University, Fort Collins, Colorado, USA}
\correspondingauthor{Senne Van Loon}{senne.van\_loon@colostate.edu}

%
%

\begin{keypoints}
\item SST-patch forcing experiments can evaluate climate emulators in an idealized setting, tracing physical relationships between variables
\item Sensitivity of global-mean radiation to local SST perturbations in AI emulator of ERA5 is qualitatively similar to physics-based models
\item Reconstruction of global-mean top-of-atmosphere radiation from historical SST anomalies fails to capture the expected negative trend
\end{keypoints}

%
%


\begin{abstract}
    The sensitivity of the radiative flux at the top of the atmosphere to surface temperature perturbations cannot be directly observed. The relationship between sea surface temperature (SST) and top-of-atmosphere radiation can be estimated with Green's function simulations by locally perturbing the sea surface temperature boundary conditions in atmospheric climate models. We perform such simulations with the Ai2 Climate Emulator (ACE), a machine learning-based emulator trained on ERA5 reanalysis data (ACE2-ERA5). This produces a sensitivity map of the top-of-atmosphere radiative response to surface warming that aligns with our physical understanding of radiative feedbacks. However, ACE2-ERA5 likely underestimates the radiative response to historical warming. We compare to two additional versions of ACE and traditional climate models. We argue that Green's function experiments can be used to evaluate the performance and limitations of machine learning-based climate emulators by examining if causal physical relationships are correctly represented and testing their capability for out-of-distribution predictions.
\end{abstract}

\section*{Plain Language Summary}
At the top of the atmosphere, the Earth's energy budget is determined by the balance between incoming and outgoing radiation. This balance is influenced by the surface temperature. The exact relationship between the two is not directly observable. Here, we estimate this relationship by utilizing a machine learning model (ACE2-ERA5) that was trained to replicate the Earth's atmosphere based on data from ERA5, a dataset that combines model predictions with observations to create a coherent picture of the atmosphere evolution since 1940. By performing idealized experiments, we investigate the relationship between surface temperature and the top-of-atmosphere radiation in ACE2-ERA5. These idealized experiments can be used to evaluate the performance of machine learning models in climate science, by testing if they correctly represent the physical relationships between variables and if they can make predictions outside of the range of their training data. Although the results are promising, ACE2-ERA5 does not succeed in predicting the top-of-atmosphere radiation based on historical warming. The idealized simulations offer a way to systematically test machine learning-based climate emulators and indicate that the emulators might not be ready yet to produce reliable predictions.

%
%

\section{Introduction}
\subsection{Motivation}

Understanding how the climate system responds to forcing is one of the most fundamental questions in climate science. To answer this question, we currently depend on general circulation models (GCMs) to correctly model the physics of the coupled ocean-atmosphere system. To date, GCMs are the best tool we have to study the climate system on long timescales, and over the course of decades we have developed an understanding of which problems can or cannot be addressed with this tool \cite{Randall18}. For example, GCMs have given us the ability to constrain future projections for many atmospheric variables, such as global-mean surface temperature, and have helped to attribute and understand human influences on the climate \cite{IPCC_2013_WGI_Ch_9,IPCC_2021_WGI}. However, these GCMs are generally computationally expensive and rely on parametrizations to represent sub-gridscale processes and radiative effects, which can lead to biases compared to observations \cite<e.g.,>[]{Soden18, ipccAR6chap7}. For instance, GCMs struggle to correctly simulate cloud processes and fail to reproduce the observed patterns of surface warming and precipitation \cite<e.g.,>[]{Xiang17,Seager19,Wills22,Rugenstein23b,Zheng23}. 

Recent advances in machine learning (ML) have made it possible to emulate the physical processes in the atmosphere in a computationally efficient way \cite<e.g.,>[]{Watt-Meyer23,Kochkov24,Guan24,Cachay24,Cresswell-Clay24,Watt-Meyer24}. These emulators replace parts or the entirety of the traditional GCMs with deep learning methods by extracting the relevant physical relationships from observational products or other climate models. The progress made over the last few years has been significant and machine learning is showing great promise to address many open problems in climate science \cite{Eyring24,Bracco25}. Yet, these ML-based emulators have to be evaluated before they can be trusted to answer actual scientific questions similar to traditional GCMs \cite{Ullrich25}.

Evaluating ML climate emulators can go hand-in-hand with increasing our scientific understanding of the climate system. Thus, the purpose of this work is twofold: (1) to propose a framework in which ML-based climate emulators can be tested in an idealized, yet process-based, setting, and (2) to evaluate the response of the top-of-atmosphere (ToA) radiation to sea surface temperature (SST) anomalies from reanalysis-based simulations. To achieve this, we perform atmospheric Green's function (GF) simulations \cite{Bloch-Johnson24} with the Ai2 Climate Emulator \cite<ACE,>[]{Watt-Meyer23,Duncan24,Watt-Meyer24}, and qualitatively compare the results to physics-based GCMs. Our results indicate that GCMs and ACE trained on reanalysis have similar sensitivities of the ToA radiation response to surface warming, but that ACE might not be ready yet to reliably predict the ToA radiation, possibly due to the lack of energy conservation in the emulator. 

\subsection{Green's function experiments with the Ai2 Climate Emulator}

In the context of this work, the Green's function (GF) refers to the linear response of the atmosphere to local perturbations of SST \cite{Branstator85,Barsugli02,Bloch-Johnson24}. Computing the GFs requires the simulation of thousands of model years, but, once calculated, can be used to study the response of any simulated variable to any SST anomaly with virtually no additional computational expense. Moreover, because the GF portrays a causal response, the resulting sensitivity maps enhance our understanding of the underlying physical mechanisms of the response \cite<e.g.,>[]{Alessi23}. Therefore, GFs can be interpreted as a ``distilled'', physically interpretable, representation of the original model \cite{Hsieh23}. 

In particular, the GF method has been successful in estimating the response of ToA radiation to SST anomalies and understanding the varying radiative feedbacks over the historical period \cite<e.g.,>[]{Zhou17,Dong19}. This problem is especially interesting because it directly influences the global energy budget at the top of the atmosphere: $N=F+R$, where $N$ is the net radiative imbalance, $F$ is the radiative forcing (due to, e.g., CO$_2$ or aerosol emissions), and $R$ is the radiative response to surface warming. The radiative imbalance $N$ is a measure of the excess energy flux into the Earth system ($N=0$ in equilibrium) and has been continuously observed by satellites since 2000 \cite{Loeb18}. To first order, $R = \lambda T_g$ is a function of surface temperature, and, in a stable climate, acts as a restoring term for the energy budget. The feedback $\lambda$ varies with the SST warming pattern, according to the so-called ``pattern effect'' \cite{Andrews15,Stevens16}. Therefore, knowing how sensitive $R$ is to varying SST is crucial to understand how much the atmosphere warms in response to a radiative forcing \cite<e.g.,>[]{Senior00,Andrews15,Andrews22,Rugenstein23}.

Importantly, only $N$ but not $R$ can be observed. As such, we rely on models to estimate $R$. GF simulations are one way of estimating the relationship between the ToA radiation and surface temperature, but -- to date -- they have always relied on GCMs, and an observation-based relationship has been elusive so far. As mentioned above, GCMs rely on parametrizations to model this relationship, which are known to have biases \cite{Soden18}. Therefore, there is an ongoing search for alternative tools to evaluate the ToA response to surface warming based on various data-driven techniques and eventually contrasting models with observations \cite<e.g.,>[]{Bloch-Johnson20,Kang23,Falasca25,Rugenstein25,VanLoon25}.

The Ai2 Climate Emulator (ACE) is a data-driven autoregressive model that has been shown to be stable on climatic timescales \cite{Watt-Meyer23,Duncan24,Watt-Meyer24}. Version 2 of ACE \cite<ACE2,>[]{Watt-Meyer24} has been trained on the ERA5 reanalysis dataset \cite{Hersbach20}, which is based on observations. When forced with historical SST and CO$_2$ concentrations, ACE2 is able to reproduce the historical data from ERA5 well \cite<see>[]{Watt-Meyer24}. ACE2 can simulate a year in a few minutes of computation time on a single GPU, such that it can be used to perform GF simulations in just a few days (compared to a few weeks for most GCM simulations, depending on the availability of compute nodes). 


\section{Data and Methods}\label{sec:methods}

\subsection{Ai2 Climate Emulator (ACE)}\label{sec:ACEmethods}

\begin{figure}
    \centering
    \hspace*{-2cm}\includegraphics{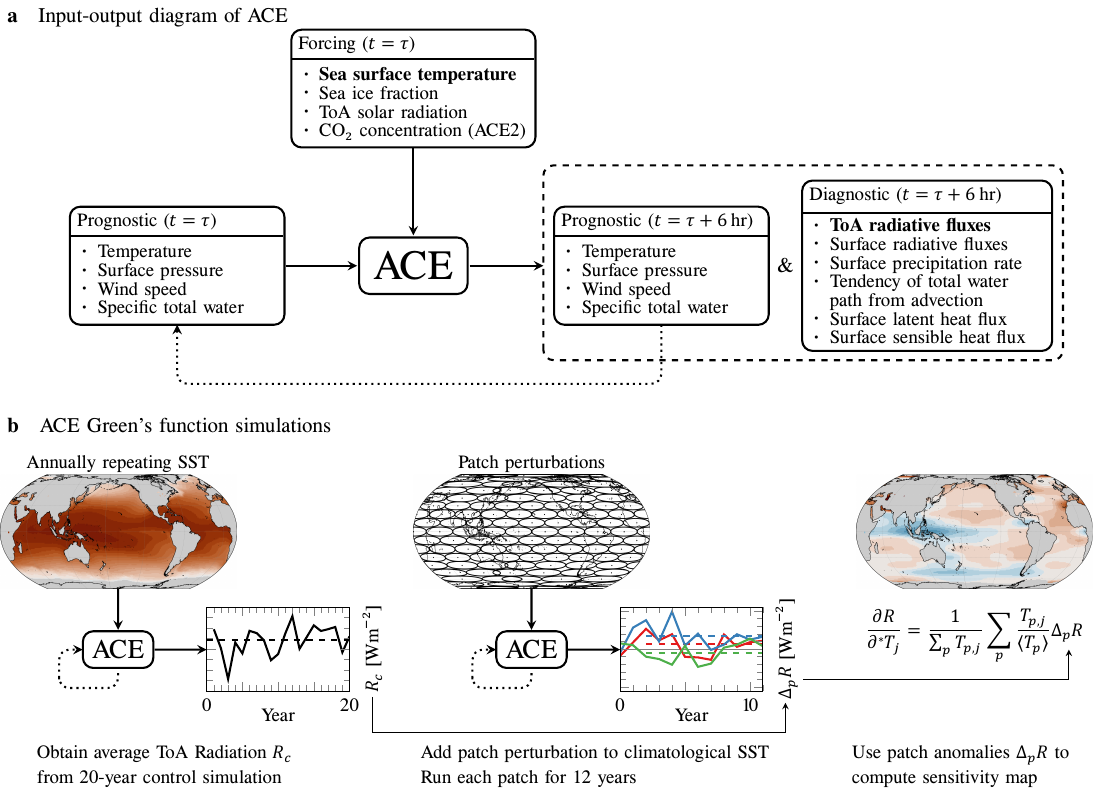}
    \caption{{\bf a} Schematic of the inputs and outputs of ACE \cite<based on>[see their Table~1]{Watt-Meyer23}. Here, we use SST forcing and ToA radiative fluxes as output. {\bf b} Schematic of Green's function (GF) simulations with ACE \cite<based on>[see their Figure~1]{Bloch-Johnson24}.
    }
    \label{fig:GFsetup}
\end{figure}

We use three already-trained versions of ACE: ACE-FV3 \cite{Watt-Meyer23}, ACE-EAM \cite{Duncan24}, and ACE2-ERA5 \cite{Watt-Meyer24}, see Supplemental~Table~S1. ACE is a neural network based on Spherical Fourier Neural Operators \cite<SFNO;>[]{Bonev23} that is designed to emulate a model of the atmosphere. All three versions of ACE are atmosphere-only autoregressive models that predict the state of the atmosphere at time $t=\tau+\SI{6}{hr}$, based on the current state ($t=\tau$) and a forcing that sets the boundary conditions. At every time $t=\tau$, ACE uses the forcing climatology (e.g., sea surface temperature) and the prognostic variables (e.g., air temperature) at $t=\tau$ to predict the state of the atmosphere six hours into the future ($t=\tau+\SI{6}{hr}$). The output state at $t=\tau+\SI{6}{hr}$ includes prognostic variables and diagnostic variables (output only, e.g., radiative fluxes). The prognostic variables at $t=\tau+\SI{6}{hr}$, together with the forcing variables, can then be used as input to predict the state at $t=\tau+\SI{12}{hr}$, and so on. Each variable is available on a $\ang{1}\times\ang{1}$ grid and eight vertical levels. We found all versions of ACE to be stable for at least four decades, in agreement with \citeA{Watt-Meyer23, Duncan24, Watt-Meyer24}. All versions of the emulator are forced by ToA incoming solar radiation (rsdt), sea-ice fraction (SIC), and sea surface temperature (SST). ACE2-ERA5 is additionally forced by CO$_2$ concentration. Fig.~\ref{fig:GFsetup}{\bf a} provides an overview of the inputs and outputs of ACE.

ACE-FV3 \cite{Watt-Meyer23} is trained on output from FV3GFS \cite{Zhou19}, the atmospheric model for the Global Forecast System (GFS) used for operational weather prediction. Training data was generated by running FV3GFS forced with an annually repeating climatology of rsdt, SIC, and SST, averaged from the period 1982-2012, keeping greenhouse gas and aerosol concentrations fixed at a 2020 value. In total, 10 ensemble members, each with 10-year-long simulations, were used to train ACE-FV3. 

ACE-EAM \cite{Duncan24} is trained on output from the E3SM Atmosphere Model (EAMv2), the atmospheric component of the Department of Energy's Energy Exascale Earth System Model \cite<E3SMv2>[]{Golaz22}. Similar to ACE-FV3, ACE-EAM was trained on an annually repeating climatology of rsdt, SIC, and SST (averaged over 2005-2014, based on observations), while emissions were kept fixed at the level of 2010. A 42-year-long simulation was used to train ACE-EAM.

ACE2-ERA5 \cite{Watt-Meyer24} is the only version we use that is trained on historical data, based on output from the fifth generation ECMWF reanalysis \cite<ERA5>[]{Hersbach20}. ERA5 data is regridded to a \ang{1} horizontal resolution and eight vertical levels. The training data consists of three periods: 1940-1995, 2011-2019, and 2021-2022, while other years are kept for validation and testing. Apart from rsdt, SIC, and SST, ACE2-ERA5 is also forced by CO$_2$ concentration.

All versions of ACE have been previously trained on different datasets and underlying atmospheric models. If ACE is a reliable emulator, it will carry over the same strengths and shortcomings of the training data. GCMs are commonly tuned to obtain realistic global-mean ToA radiative fluxes, enforcing energetic constrains \cite{Hourdin17,Wild20}. For example, in the coupled GCM E3SM, the atmospheric component EAMv2 is tuned to obtain a near-zero ToA radiative flux in a long pre-industrial control simulation \cite{Golaz22}. Therefore, we might expect ACE-EAM to carry over this energy constraint. In contrast, numerical weather prediction models like FV3GFS are not tuned to reproduce the global ToA radiation, as they are not necessarily built for predictions on climate timescales, which require energy constraints. ERA5 is constrained by assimilating observations, but does not assimilate observations of the ToA radiative fluxes \cite{Hersbach20}. \citeA{Wild24} showed that ERA5 performs reasonably well in capturing the long-term average and internal variability of the net ToA radiation, but does not reproduce the historical trend of the global-mean energy imbalance. Although physical constraints can be imposed on ACE2, ACE2-ERA5 only conserves the global mean moisture budget, not the energy budget \cite{Watt-Meyer24}.

\subsection{Green's function experiments}\label{sec:GFmethod}

The Green's function (GF) method is outlined in Fig.~\ref{fig:GFsetup}{\bf b}. We generally follow the protocol as outlined by the Green's Function Model Intercomparison Project \cite<GFMIP,>[]{Bloch-Johnson24}, although we make minor changes when applying it to ACE. The GFMIP protocol in \citeA{Bloch-Johnson24} is designed to investigate differences across GCMs by standardizing the simulations. We focus on the GF for the globally averaged net ToA radiation $R$, defined as the sum of incoming solar radiation, reflected solar radiation, and outgoing longwave radiation at the top of the atmosphere (ToA; positive downwards). However, we note that the same protocol can be applied to any atmospheric variable that is output by ACE, without performing additional model runs. In the following, we refer to the ``Green's function'' or GF as the partial derivative of $R$ with respect to local SST, but note that the GF method can be more generally applied to different simulated variables.

First, we run a 20-year control simulation for each version of ACE forced with an annually repeating climatology. For ACE-FV3 and ACE-EAM, we use the climatology it was trained on, while for ACE2-ERA5 we compute the ERA5 climatology from 1971-2020. The climatology is calculated by averaging the values in all years 1971-2020 for every grid point and 6-hourly time step (CO$_2$ is held constant at its average 1971-2020 value). The 20-year-mean, globally averaged, ToA radiation of the control simulation, $R_c$ (dashed black line in Fig.~\ref{fig:GFsetup}{\bf b}), is used as the reference state for the GF simulations.

Next, we perform patch simulations, each with the same initial conditions taken from the end of the control run. We use the same annually repeating climatology, but add a local perturbation to the SST:
\begin{equation}
    T_p(\varphi,\vartheta) = \left\{
    \begin{aligned}
        &A 
        \cos^2 \left( \pi \frac{\varphi-\varphi_p}{\delta\varphi_p} \right)
        \cos^2 \left( \pi \frac{\vartheta-\vartheta_p}{\delta\vartheta_p} \right) 
        && \left\{\begin{aligned}
            \varphi - \varphi_p &\in [-\delta\varphi_p,\delta\varphi_p]\\
            \vartheta - \vartheta_p &\in [-\delta\vartheta_p,\delta\vartheta_p]\\
        \end{aligned} \right.\\
        & 0 && \kern1em\text{elsewhere}
    \end{aligned}
    \right. .
\end{equation}
Here, $A$ is the amplitude of the perturbation, $(\varphi,\vartheta)$ are latitude and longitude, $(\varphi_p,\vartheta_p)$ denotes the center of the perturbation, and $(\delta\varphi_p,\delta\vartheta_p)$ are the latitudinal and longitudinal widths of the perturbation. We use the patch layout of \citeA<>[see their Table 1]{Bloch-Johnson24}; a sketch of the different patches is shown in Fig.~\ref{fig:GFsetup}{\bf b}, where for each patch the contour at $A/2$ is shown. In total, there are 109 patches that cover the entire globe. We perform simulations for positive ($A=+\SI{2}{K}$) and negative ($A=-\SI{2}{K}$) perturbations, with each patch run for 12 years, totalling to 2616 simulation years. The ToA radiation of each patch simulation is averaged over the last 10 years, allowing for two years of spin-up. 

Finally, the difference $\Delta_p R= R_p-R_c$ between the patch radiation $R_p$ and the control $R_c$ (colored dashed lines in Fig.~\ref{fig:GFsetup}{\bf b}) is used to calculate the GF. At each grid-box $j = (\varphi_j,\vartheta_j)$, the GF of the ToA radiation $R$ is given by
\begin{equation}
    \frac{\partial R}{\partial^\ast T_{j}} 
    = \frac{
        1
    }{
        \sum_p T_{p,j}
    }
    \sum_p \frac{T_{p,j}}{\langle T_p \rangle} \Delta_p R
    \label{eq:GF}
\end{equation}
Here, $\langle T_p \rangle = \sum_j a_j T_{p,j}$, the weighted average of the patch temperature, $T_{p,j}=T_p(\varphi_j,\vartheta_j)$, and $a_j$ the ice-free ocean area of grid-box $j$ normalized to the total ice-free ocean area. The asterisk in Eq.~\eqref{eq:GF} denotes the area-normalized derivative, related to the gradient as ${\partial R}/{\partial T_{j}} = a_j \, {\partial R}/{\partial^\ast T_{j}} $. The normalized derivative makes the quantity independent of grid-box area and allows for a direct comparison between models with different spatial resolution \cite{Bloch-Johnson24}. A separate GF is calculated for positive and negative patch perturbations, and the average of the two is taken as the final GF.

We compare the GFs of ACE to GFs of five GCMs that were calculated from the average of positive and negative patch simulations: CAM5 \cite{Neale10,Zhou17}, CanESM5 \cite{Swart19}, ECHAM6 \cite{Stevens13,Alessi23}, GFDL-AM4 \cite{Zhao18,Zhang23}, and HadAM3 \cite{Pope00}. These all differ slightly in their patch setup and control climatology, but use a similar protocol as described above \cite<see Figure 2 in>[]{Bloch-Johnson24}. GF simulations of the GCMs ACE has been trained on (FV3GFS and EAMv2) are not yet available, hence, we do not have a ``ground truth'' to compare our simulations to. Therefore, we only qualitatively compare the GFs of the traditional, physics-based GCMs to the ones obtained with different ACE versions. The GCM GFs are all scaled to a common \ang{3.75}$\times$\ang{2.5} grid (the lowest resolution among the GCMs) using a conservative method with periodic boundary conditions.

\section{Results}\label{sec:Results}

\subsection{ACE Green's function sensitivity maps}\label{sec:ACEGF}

\begin{figure}
    \centering
    \includegraphics{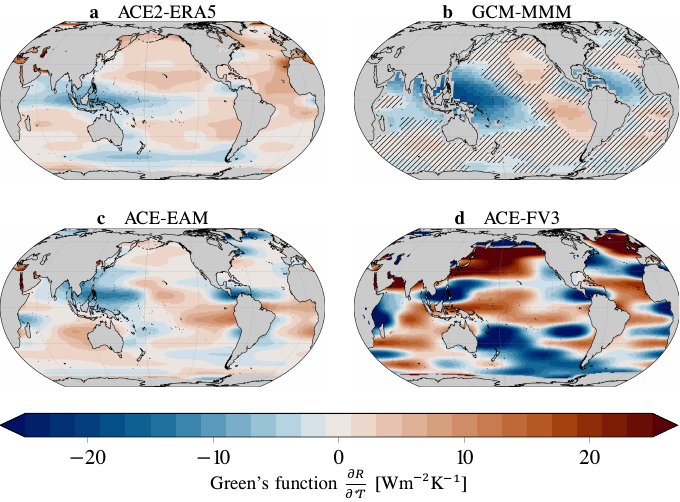}
    \caption{Green's functions of ({\bf a}) ACE2-ERA5, ({\bf b}) Multi-model mean (MMM) of 5 GCMs \cite{Bloch-Johnson24}, ({\bf c}) ACE-EAM, and ({\bf d}) ACE-FV3. Note that the the resolution of ACE is \ang{1}$\times$\ang{1}, while the GCMs are on a \ang{3.75}$\times$\ang{2.5} grid. Hatching in {\bf b} indicates regions where at least one GCM disagrees on the sign of the Green's function.}
    \label{fig:GF}
\end{figure}

In Fig.~\ref{fig:GF}, we compare the GFs of ACE-FV3, ACE-EAM, and ACE2-ERA5 with the average GF across the five GCMs. The sensitivity maps visualize how the global-mean ToA radiation $R$ responds when increasing the SST in a single grid-box. Thus, they represent the sum of local and remote radiative feedbacks to changing SST. The reanalysis-based GF of ACE2-ERA5 (Fig.\ref{fig:GF}{\bf a}) is qualitatively similar to the GCM results (Fig.~\ref{fig:GF}{\bf b}), although the amplitude of the local feedbacks is lower. Similarly, ACE-EAM (Fig.~\ref{fig:GF}{\bf c}) finds the general structure of the GCM GF. In contrast, ACE-FV3 (Fig.~\ref{fig:GF}{\bf d}) shows a much noisier map. 

Fig.~\ref{fig:GF}{\bf a} depicts the first GF based on reanalysis data. As explained in section~\ref{sec:ACEmethods}, ERA5 constrains its predictions by assimilating observations, but does not use direct observations of the ToA radiative imbalance \cite{Hersbach20}. The fact that ACE2-ERA5 is able to reproduce the main features of the GCM GF (Fig.~\ref{fig:GF}{\bf a}~versus~{\bf b}) is not trivial, and indicates that ACE2-ERA5 captures the main physical mechanisms of the radiative feedbacks. Still, there are regional differences between the ACE2-ERA5 and GCM GFs and the amplitude of the local feedbacks is lower in ACE2-ERA5. We emphasize that the \emph{true} sensitivity map is unknown, and thus there is no a priori reason to trust the GCMs over ACE2-ERA5.

The ACE2-ERA5 GF shows features that align with our physical understanding of local and remote feedbacks. We expect to see negative feedbacks in regions of deep convection, such as the tropical West Pacific, and positive feedbacks in regions of subsidence, such as the subtropical East Pacific. This behavior is mostly driven by the sensitivity of shallow marine stratocumulus clouds to local and remote changes in SST \cite{Zhou16,Andrews18,Myers21,Myers23,Rugenstein23}. For example, local warming in the region of the low cloud decks in the subtropical East Pacific leads to a decrease in low-level clouds, which in turn decreases the outgoing shortwave radiation (positive feedback, Supplemental~Figure~S1). In contrast, warming the surface in the West Pacific warm pool leads to warmer upper troposphere in the subtropical East Pacific due to teleconnections. This warming stabilizes the air in regions with low clouds, increasing the amount of clouds and therefore the outgoing shortwave radiation (negative feedback, Supplemental~Figure~S2).

GCMs capture the same physical mechanisms of feedbacks: they agree on the sign of the GF in regions of deep convection and subsidence (Fig.~\ref{fig:GF}{\bf b}). The ACE-EAM GF (Fig.~\ref{fig:GF}{\bf c}) has a similar structure, indicating that ACE-EAM has learned the relationship between SST and $R$ from EAMv2. Still, since ACE-EAM was trained on an annually repeating climatology, there is no inherent reason why this relationship would be preserved when forcing it with patch perturbations. Even if a $\pm\SI{2}{K}$ perturbation falls within the local seasonal cycle of SST, at some time of the year the perturbed SST will be outside the range of the training data. This is especially true for the tropics, where the seasonal cycle is small, and large regions never experience variations larger than $\pm\SI{2}{K}$ (Supplemental~Figure~S3{\bf a}). This means that the GF simulations are truly out-of-distribution for ACE-EAM, pointing to the possibility of using ACE to investigate future climates.

In contrast, ACE-FV3 fails to capture the expected relationship between SST and $R$ (Fig.~\ref{fig:GF}{\bf d}). This might be because numerical weather prediction models like FV3GFS (on which ACE-FV3 was trained) are not energetically closed, or because of differences in the training setup compared to other versions of ACE. Similar to ACE-EAM, the GF simulations sample ACE-FV3 outside of its trained domain (Supplemental~Figure~S3{\bf b}), but reducing the patch amplitude to $A=\pm\SI{1}{K}$ did not result in a more physical sensitivity map (Supplemental~Figure~S4). The ACE-FV3 GF shows a pattern similar to the patch layout (Fig.~\ref{fig:GFsetup}{\bf b}), indicating that responses can vary considerably between adjacent patch perturbations. ACE-FV3 tends to overestimate the variability of $R$ on yearly timescales (Supplemental~Figure~S6), such that longer temporal averages might be necessary to filter out the noise. However, lengthening the patch simulations to 30 years did not improve the results.

\subsection{Historical ToA radiative response to surface warming}\label{sec:ACE_pred}

\begin{figure}
    \centering
    \includegraphics{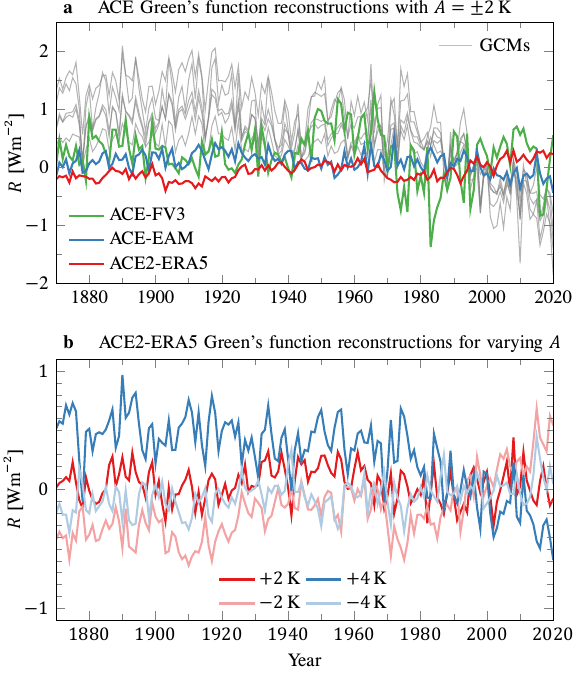}
    \caption{Historical reconstruction of the net ToA radiation $R$, formed by convolving the Green's functions with AMIP sea surface temperature anomalies with respect to 1971-2020.}
    \label{fig:historical}
\end{figure}

GFs not only give us physical insight into local feedbacks, but they can be used to estimate the radiative response to any SST forcing. The GFs depict the partial derivative of $R$ with respect to local SST ($T_j$), such that in a linear approximation
\begin{equation}
    R = \sum_j a_j \, \frac{\partial R}{\partial^\ast T_{j}} T_j + \mathcal{O}(T^2).
    \label{eq:R}
\end{equation}
That is, given any SST map, the ToA radiation can be approximated by using equation~\eqref{eq:R}. To test whether or not the ACE GFs can be used for this purpose, we convolve the GF with SST maps from the AMIP dataset \cite{Taylor00}, which contains observed SST from 1870-2020. The resulting time series of $R$ is shown in Fig.~\ref{fig:historical}. Each line was calculated by conservatively regridding the AMIP SST to ACE's native grid, multiplying it with the GF, and averaging the result weighted by the ice-free ocean area. As a reference, we show the predictions from the GCM GFs as gray lines calculated in the same way as the ACE GF predictions. 

Although we do not know the absolute truth for the historical $R$, we expect it to become more negative in a warming world. That is, with increasing global-mean temperature, we expect a greater outgoing energy flux to counteract the forcing. This decline is not predicted by the ACE2-ERA5 GF (Fig.~\ref{fig:historical}, red line). Moreover, the variability of $R$ is much smaller than in the observed ToA radiative imbalance (Supplemental~Figure~S7). Similarly, the ACE-EAM GF does not predict a realistic $R$, although it predicts a slight decrease in the last two decades. The annual variability in the predicted $R$ from ACE-FV3 is larger, but shows unrealistic decadal variations.

Even though the ACE2-ERA5 sensitivity map looks similar to the GCM GFs (Fig.~\ref{fig:GF}), its predicted ToA radiative response $R$ does not (Fig.~\ref{fig:historical}). This could be due to ACE2-ERA5 not capturing the full response of $R$ to SST, the GF protocol, or both. To test the former, we run ACE2-ERA5 forced with historical SST from ERA5, and compare the predicted ToA radiation with the ``truth'' from ERA5 (Supplemental~Figure~S7). The output of this historical run reflects how well ACE2-ERA5 performs in simulating the ToA radiative imbalance as compared to the underlying dataset (ERA5 reanalysis). The ERA5 data itself does not capture the correct trend in the ToA radiation, but performs well on its variability, as compared to direct observations \cite<CERES-EBAF,>[]{Loeb18}. Moreover, ACE2-ERA5 is not able to reproduce ERA5 ToA radiation on yearly timescales, indicating that ACE2-ERA5 misses some of the response of $R$ to SST. Finally, ACE2 exhibits nonphysical behavior when forced with historical SST but keeping CO$_2$ fixed \cite{Watt-Meyer24}, which can lead to unrealistic patch responses if ACE2 does not correctly separate CO$_2$ and SST changes.

To further test the GF setup, we calculate the ACE2-ERA5 GF for $A=\pm\SI{4}{K}$ (Supplemental~Figure~S5). GFs for $A=\pm\SI{2}{K}$ and $A=\pm\SI{4}{K}$ are qualitatively similar, indicating that the ACE2-ERA5 GF is somewhat robust to the amplitude of the patches. However, the predicted $R$ is very sensitive to the patch amplitude and its sign. In ACE2-ERA5, using the GF with $A=+\SI{4}{K}$ (positive patch perturbations only), the reconstructed $R$ looks more realistic (dark blue line in Fig.~\ref{fig:historical}{\bf b}) than the $A=\pm\SI{2}{K}$ reconstruction (red line in Fig.~\ref{fig:historical}{\bf a}). 

Generally, GF predictions are highly nonlinear in the sign of $A$ \cite{Williams23,Bloch-Johnson24}. This is a robust feature in GCMs, but not well understood, and ACE captures the same property (Supplemental~Figure~S8), indicating that it might be a property of the climate system. Using $A>0$ only, the GFs mostly predict a decrease in $R$ over the historical period, while using $A<0$ only, the GF mostly predict an increase in $R$. This is true for all versions of ACE (apart from ACE2-ERA5 with $A=+\SI{2}{K}$) and for GCMs. Surprisingly, the reconstructed $R$ from ACE-FV3 is comparable to $R$ based on the GCM GFs, even though its GF did not follow our expectations (Fig.~\ref{fig:GF}{\bf d}). This is likely due to competing biases in the ACE-FV3 GF, which cancel out in the reconstruction. These results point to some of the limitations of the GF method, discussed in more detail in the following section.

\section{Discussion}\label{sec:discussion}

Estimating the ToA radiative response to surface warming is a topic of debate, and the true relationship between $R$ and SST is unknown \cite<e.g.,>[]{Sherwood20,Rugenstein23}. To study this problem, we have thus far had to rely on GCMs because the observational record is too short to find a robust relationship between the two. Even if a longer record existed, we can only directly observe the top-of-atmosphere radiative imbalance $N=F+R$ and not $R$ itself. Because we cannot turn off or measure the forcing $F$ in the real world, we only have access to the internal variability of the ToA fluxes \cite<which are assumed to be similar for $N$ and $R$, e.g.,>[]{Dessler18}. Thus, when using observations to constrain future projections, we need to assume that internal variability of $R$ also explains its forced response \cite{Rugenstein25}. 

Green's function (GF) experiments with ACE2-ERA5 allow us to overcome these issues. Because the GFs are calculated by only changing the SST, all other forcing agents (e.g., CO$_2$ or aerosols) are essentially kept fixed, such that $F=0$ and $N=R$. This is similar to performing AMIP experiments in GCMs \cite{Webb17}. However, these experiments are costly to run, have their own model-specific biases and unknowns in the physics, and require coordinated protocol efforts, for which the development cycle usually lags behind observations. Thus, if we are able to trust the ACE2-ERA5 GF, this method offers a new way to remove the radiative forcing from the \emph{observed} climate system. That is, the ACE2-ERA5 GF represents the historical, reanalysis-based, climate with $F=0$. 

As we have shown, however, ACE2-ERA5 may not yet be ready for this purpose. ERA5 comes with its own biases with respect to observations \cite<e.g.,>[]{Wild24,Dussin25}. These biases are propagated to ACE2-ERA5, even if the emulator would exactly replicate the behavior of ERA5 (which it does not, e.g., Supplemental~Figure~S7). Overall, machine-learning methods are known to perform poorly on out-of-distribution tasks. Because ACE2-ERA5 is trained on historical reanalysis data, it is not expected to perform well in far out-of-distribution scenarios, such as quadrupling of CO$_2$ concentration or a much warmer world \cite{Watt-Meyer24}. Given this, the fact that ACE2-ERA5 is able to reproduce the GCM GF qualitatively is a big achievement, and promising for further development of climate emulators. Even for patch perturbations of larger amplitude ($A=\pm\SI{4}{K}$), the ACE2-ERA5 GFs remain realistic (Supplemental~Figure~S5). ACE2 allows for the inclusion of physical constraints, which was used to conserve the global mean moisture budget in ACE2-ERA5 \cite{Watt-Meyer24}. Using this feature to constrain the global energy budget would be a possible way forward to better represent the ToA radiative response to surface warming. We hypothesize that a closed energy budget is a necessary condition for climate emulators to accurately make predictions.

Besides the GF method, other approaches have been used to study the spatial relationship between $R$ and SST \cite<e.g.,>[]{Bloch-Johnson20,Falasca25,Rugenstein25,VanLoon25}. These different methods generally agree on the overall structure of the sensitivity maps (e.g., Fig.~\ref{fig:GF}{\bf a}-{\bf c}), suggesting that this is a robust feature across models. However, approaches differ in the specifics of the sensitivity maps, such as the importance of small-scale features or the amplitude of the local feedbacks, while still being able to reproduce $R$ with similar skill \cite{Rugenstein25}. Because ACE2-ERA5 is computationally efficient, we can use it to investigate these differences once we trust its predictions, for example, by running different patch setups and amplitudes or multiple patches at once to investigate nonlinearities.

By performing GF simulations, the sensitivity of climate emulators can be tested. \citeA{Ullrich25} called for systematic testing of machine learning-based Earth system models in order to demonstrate their use for scientific discovery in the same manner as physics-based GCMs. The GF method could be one of these tests, as it probes how well the emulator performs in an idealized setting outside of its trained domain. To fully trust emulators, this method should be used in conjunction with other metrics, such as forced climate change experiments, the capability to reproduce internal variability of the climate system, specific feedbacks, etc. The GF sensitivity maps offer initial insight into whether or not the emulator responds realistically to variations in SST, and can be directly compared to GCMs. By looking at the response to individual patch perturbations (e.g., Supplemental~Figures~S1~and~S2), it is possible to investigate causal relationships between SST and other variables. In this work, we only provide a qualitative comparison of ACE and GCMs, but a quantitative comparison could be performed with an emulator trained on a GCM for which GF simulations are available. 

However, GF simulations come with its own limitations. Even across different GCMs, there are large variations in the representation of the sensitivities, and the absolute truth is not known \cite{Bloch-Johnson24}. Although we have some theories about the physical processes that govern positive or negative feedbacks in the sensitivity maps (e.g., low cloud feedbacks in marine stratocumulus regions), we do not know the exact extent, spatial detail, or amplitude of the local feedbacks \cite{Myers21,Myers23,Ceppi24}. Aside from the model used, the GFs depend on choices of the exact patch setup or length of the simulations and do not take into account nonlinearities of the climate system \cite{Bloch-Johnson21,Williams23}. For example, the reconstructed $R$ depends on the amplitude of the patch perturbation (Fig.~\ref{fig:historical}{\bf b} and Supplemental~Figure~S8). This makes comparing sensitivity maps between emulators more of a qualitative test than a quantitative assessment. 

In summary, we have performed Green's function simulation with different versions of the Ai2 Climate Emulator (ACE). The Green's function of ACE2-ERA5, which was trained on reanalysis data, shows promising qualitative agreement with expectations, but cannot recreate a realistic top-of-atmosphere radiative response to surface warming in the historical period. We conclude that Green's function experiments offer a way to test climate emulators in an idealized setting and investigate their sensitivity to external forcing. Such simulations could be part of a broader testbed for machine learning-based atmospheric models to assess their viability for scientific discovery.

%
%

\section*{Open Research Section}
All ACE code, checkpoints, and forcing data are freely available through the Ai2 Hugging Face repository at \url{https://huggingface.co/collections/allenai/ace-67327d822f0f0d8e0e5e6ca4}. The code used to perform the GF simulations is available at \url{https://github.com/SnnVL/GF4ACE} and output data is available at \url{https://doi.org/10.5061/dryad.d2547d8cf} \cite{VanLoon25_GF4ACE}. The GFs from CAM5, CanESM5, ECHAM6, GFDL-AM4, and HadAM3 were obtained from \citeA{Bloch-Johnson23}. Monthly historical temperature (AMIP) was downloaded from \url{https://gfmip.org}.

\acknowledgments
This work was supported, in part, by the Regional and Global Model Analysis program area of the U.S. Department of Energy's (DOE) Office of Biological and Environmental Research (BER) as part of the Program for Climate Model Diagnosis and Intercomparison project. EAB was also supported by a grant from the Heising-Simons Foundation.

\clearpage
\section*{Supporting Information}
\setcounter{figure}{0}
\setcounter{table}{0}
\renewcommand{\figurename}{{\bf Supporting Information Fig.}}
\renewcommand{\tablename}{{\bf Supporting Information Table}}
\renewcommand{\thefigure}{S\arabic{figure}}
\renewcommand{\thetable}{S\arabic{table}}

\begin{figure}[h]
    \centering
    \includegraphics{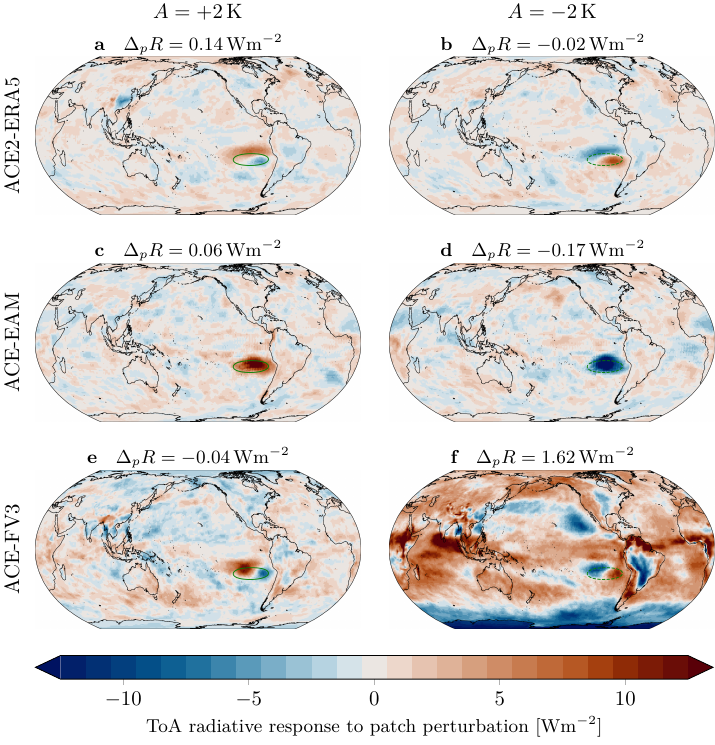}
    \caption{Radiative response to a patch perturbation in the subtropical East Pacific. Left column shows a positive SST perturbation ($A=\SI{2}{K}$) and right column a negative SST perturbation ($A=-\SI{2}{K}$), for ACE2-ERA5 ({\bf a}-{\bf b}), ACE-EAM ({\bf c}-{\bf d}), and ACE-FV3 ({\bf e}-{\bf f}). The values above each map show the global-mean values of the response, used to calculate the Green's function.
    }
    \label{patch_response_SEP}
\end{figure}

\begin{figure}[h]
    \centering
    \includegraphics{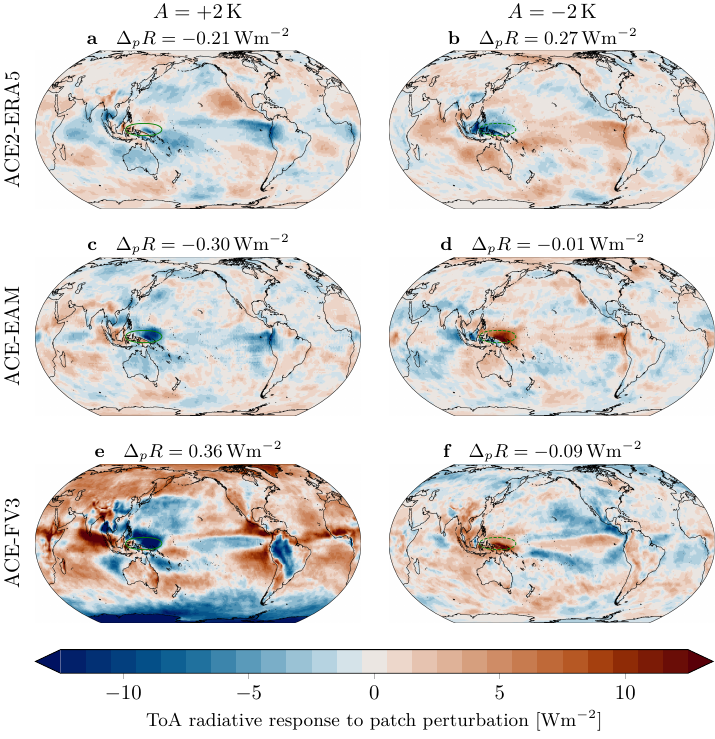}
    \caption{Radiative response to a patch perturbation in the West Pacific warm pool. Left column shows a positive SST perturbation ($A=\SI{2}{K}$) and right column a negative SST perturbation ($A=-\SI{2}{K}$), for ACE2-ERA5 ({\bf a}-{\bf b}), ACE-EAM ({\bf c}-{\bf d}), and ACE-FV3 ({\bf e}-{\bf f}). The values above each map show the global-mean values of the response, used to calculate the Green's function.
    }
    \label{patch_response_EWP}
\end{figure}

\begin{figure}[h]
    \centering
    \includegraphics{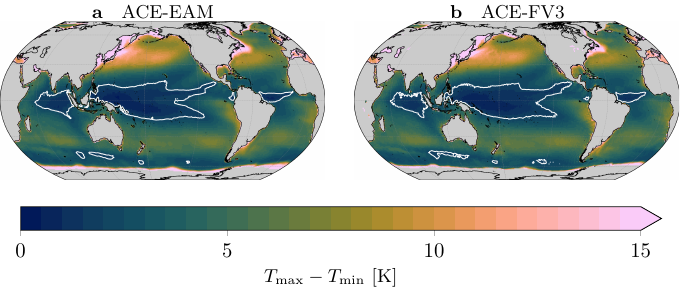}
    \caption{Amplitude of the seasonal cycle of the climatology used to train ACE-EAM ({\bf a}) and ACE-FV3 ({\bf b}). The amplitude is calculated by taking the difference between the maximum and minimum of the six-hourly data in the annual climatology. The white contour demarcates the region where the amplitude is smaller than $\SI{2}{K}$, the amplitude of the patch perturbations used in the Green's function simulations.}
    \label{SeasonalCycle}
\end{figure}

\begin{figure}[h]
    \centering
    \includegraphics{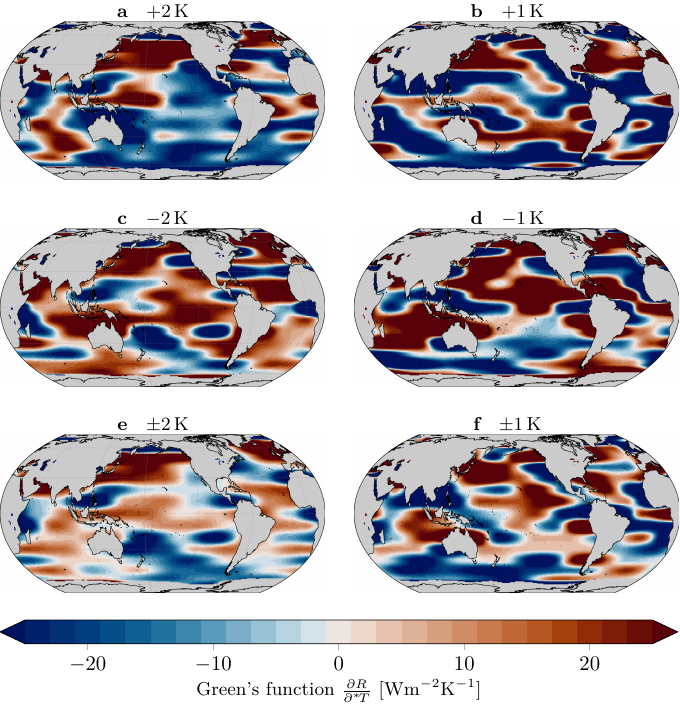}
    \caption{Green's functions of ACE-FV3 for different patch amplitudes $A$. The top row ({\bf a}-{\bf b}) shows positive patches, the middle row ({\bf c}-{\bf d}) negative patches, and the bottom row ({\bf e}-{\bf f}) the average of the positive and negative patch Green's functions.
    }
    \label{GF_fv3_T_sensitivity}
\end{figure}

\begin{figure}[h]
    \centering
    \includegraphics{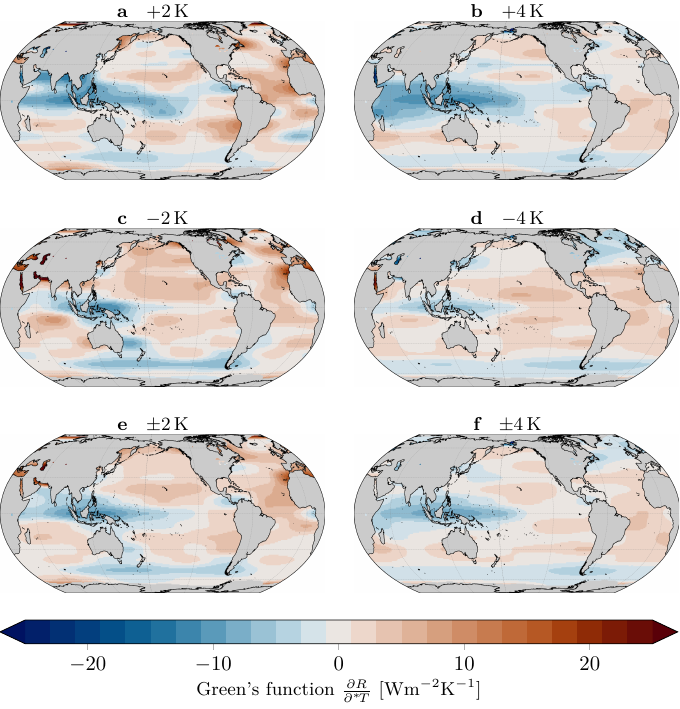}
    \caption{Green's functions of ACE2-ERA5 for different patch amplitudes $A$. The top row ({\bf a}-{\bf b}) shows positive patches, the middle row ({\bf c}-{\bf d}) negative patches, and the bottom row ({\bf e}-{\bf f}) the average of the positive and negative patch Green's functions.
    }
    \label{GF_era5_T_sensitivity}
\end{figure}

\begin{figure}[h]
    \centering
    \includegraphics{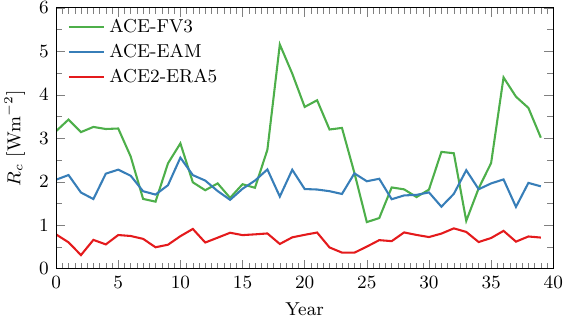}
    \caption{Top-of-atmosphere (ToA) radiation of the control run for three different versions of ACE, forced by an annually repeating climatology (see Table~\ref{tab:ACE}). Only the first twenty years are used to construct the Green's function.}
    \label{control_R}
\end{figure}

\begin{figure}[h]
    \centering
    \includegraphics{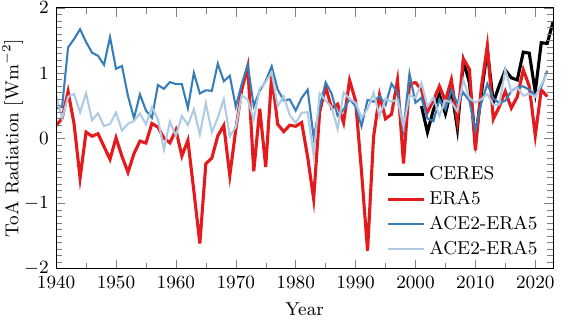}
    \caption{Top-of-atmosphere (ToA) radiation in ERA5 (red line) compared to two historical runs in ACE2-ERA5 (blue lines), starting from different initial conditions but using the same forcing. The black lines shows the observed ToA radiation from CERES-EBAF \cite{Loeb18}. The blue lines are calculated by forcing the ACE2-ERA5 model with ERA5 data from 1940-2022 for the incoming solar radiation, sea ice concentration, sea surface temperature, and CO$_2$ concentration. Because historical data is used, here the ToA radiation contains a forced signal, and is thus showing the ToA radiative imbalance $N = F + R$, with $F$ the radiative forcing.
    }
    \label{N_historical}
\end{figure}

\begin{figure}[h]
    \centering
    \hspace*{-2cm}\includegraphics{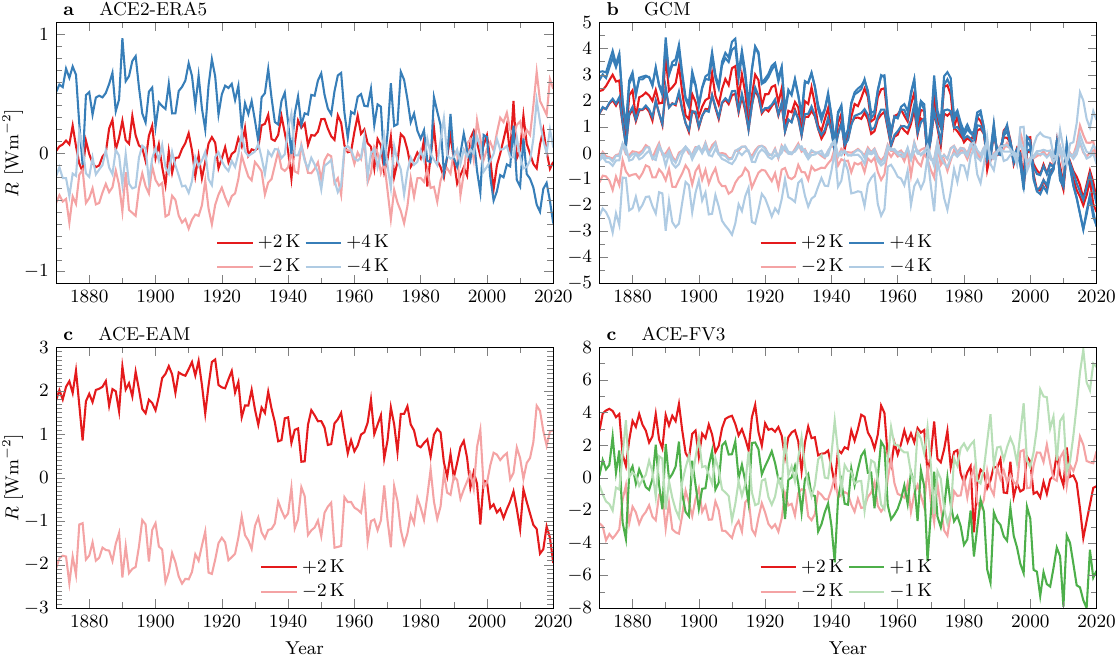}
    \caption{Historical reconstruction of the net ToA radiation $R$, formed by convolving the Green's functions computed from different patch amplitudes $A$ with AMIP sea surface temperature anomalies with respect to 1971-2020. Panels {\bf a}, {\bf c}, and {\bf d} show the reconstructions from different versions of ACE (ACE2-ERA5, ACE-EAM, and ACE-FV3, respectively). In panel {\bf b}, reconstructions from the GCM Green's functions are shown \cite{Bloch-Johnson23}. Every line represents a different GCM; the blue lines show GCMs that used $A=\pm\SI{4}{K}$, while the red lines show the GCMs that used $A=\pm\SI{2}{K}$.
    }
    \label{historical_T_sensitivity}
\end{figure}

\clearpage
\begin{table}
    \centering
    \caption{Overview of different ACE checkpoints and Green's function forcing data.}
    \begin{tabular}{r | c c | c }
        Name & Training model & Training data & control climatology \\ \hline
        ACE-FV3$^{a}$ & FV3GFS$^{b}$ & 1982-2012 climatology & 1982-2012 \\
        ACE-EAM$^{c}$ & EAMv2$^{d}$ & 2005-2014 climatology & 2005-2014 \\
        ACE2-ERA5$^{e}$ & ERA5$^{f}$ & 1940-1995, 2011-2019, 2021-2022 & 1971-2020 
    \end{tabular}
    \begin{flushleft}
        $^{a}$ \citeA{Watt-Meyer23}. $^{b}$ \citeA{Zhou19}. $^{c}$ \citeA{Duncan24}. $^{d}$ \citeA{Golaz22}. $^{e}$ \citeA{Watt-Meyer24}. $^{f}$ \citeA{Hersbach20}.
    \end{flushleft}
    \label{tab:ACE}
\end{table}

%
%

\end{document}